\documentclass[a4paper,11pt]{article}
\usepackage{nfraconf,graphicx,cite}

\usepackage{times}

\def\hub{$H_0=100$ km s$^{-1}$ Mpc$^{-1}$, $q_0 = 0.5$}

\def\sprop{$S_{\nu} \propto \nu^{\alpha}$}
\def\deg{\ifmmode $\setbox0=\hbox{$^{\circ}$}$^{\,\circ}
          \else    \setbox0=\hbox{$^{\circ}$}$^{\,\circ}$\fi\,}

\begin{document}
\baselineskip=13pt

\title{SKA in VLBI: Impact on Studies of Small Scale Structures in Active Galactic Nuclei}

\author{T.P. Krichbaum, A. Witzel, and J.A. Zensus}
\address{Max-Planck-Institut f\"ur Radioastronomie\\
Auf dem H\"ugel 69 \\
D-53121 Bonn \\
Germany \\ 
E-mail: tkrichbaum@mpifr-bonn.mpg.de}

\maketitle

\abstract{
We discuss the advantages of an
electronically steerable interferometer of a million square meters collecting area
operating at cm-wavelengths as a phased array and in conjunction with other antennas 
in VLBI mode. With a sensitivity to $\mu$Jy-level flux densities, a brightness temperature 
limit of $T_B = 10^4$ K, and an expected dynamic range in 
future VLBI images of better than $10^6$, the impact of SKA on the research on compact galactic and
extragalactic radio sources will be virtually unpredictable, but certainly
strong. In conjunction with future
space VLBI missions and complementary to planned large 
interferometers operating at millimeter wavelengths, 
SKA bears the potential to revolutionize astronomy
and our picture of the Universe. In this article we briefly outline 
the improvements expected from the use of SKA for VLBI in general. Then more
specifically, we discuss the anticipated progress for the research
on active galactic nuclei (AGN) and their powerful jets, emanating
from hypothetical supermassive black holes. We will point out that
beside a wealth of new details which SKA should reveal on jets, accretion disks and black holes,
it will help to understand the relationship between the various types of
active galactic nuclei (quasars \& galaxies) and their cosmological evolution.
}

\section{Introduction}

Since the late 1960s, the observational technique of \underline{V}ery \underline{L}ong 
\underline{B}aseline \underline{I}nterferometry (VLBI) has evolved 
enormously, yielding results from first detections of ultracompact structures and superluminal
motion in AGN in the early years to present day high fidelity monitoring 
of extragalactic radio jets, showing details at multiple frequencies of structure and kinematics
from sub-milliarcsecond to sub-arcsecond scales. Although
the use of more homogeneous VLBI arrays (e.g. with the VLBA of the NRAO, 10 identical 25\,m antennas) 
and the combination
of up to 20 antennas in so called `world array' experiments have considerably improved the
quality of the images achieved, VLBI to date still is limited 
in sensitivity (to mJy level) and also by a somewhat uneven coverage of the uv-plane, which 
adversely affects the image reconstruction. 

The somewhat stochastic distribution of radio telescopes around the world
(most of them with diameters in the 10-30 m range located on the northern hemisphere) 
causes problems with signal detection on less-sensitive baselines, gaps in the uv-plane, 
and problems with image deconvolution and sidelobes. So far, only a few of the VLBI
maps presented in the literature reach the thermal noise limit and have a dynamic
range (defined as ratio of peak flux to noise level) better than $10^4$. 
The sensitivity to partially resolved sources or source components of low surface brightness
is at present limited to a typical brightness temperature of $T_B \geq 10^{6...7}$\,K.

With system temperatures of only a few tens of Kelvin at cm-wavelengths,
and observing bandwidths already reaching the GHz range in the foreseeable future (256 MHz
to date), a further improvement of VLBI imaging capabilities can only be achieved by an
increase of the total collecting area and an improvement of the uv-coverage, i.e. increasing 
the collecting area and number of the participating antennas. 

In this contribution we shall discuss the substantial improvements to science, and 
in particular to the research on compact extragalactic radio sources,
if the planned {\underline{S}quare {\underline{K}ilometer {\underline{A}rray 
(SKA) is used in its configuration
with long baselines ($500-1000$ km maximum antenna spacing). As
stand-alone instrument its resolution will be limited to a few ten milliarcseconds,
so we assume here that SKA is added as a phased array into existing or future global 
VLBI networks operating from ground and space.

\section{Sensitivity}

For a Very Long Baseline Interferometer the 1\,$\sigma$ detection threshold $\sigma_{ij}$
(in [Jy]) for detecting the fringe visibility on a single
baseline is:

\begin{equation}
\sigma_{ij} =  \frac{1}{\eta_c} \cdot 
\sqrt{ \frac{T_{\rm sys}^{~i}}{g_i} \cdot \frac{T_{\rm sys}^{~j}}{g_j}}
\cdot \frac{1}{\sqrt{\Delta \nu \cdot \tau}}
\end{equation}

where $T_{\rm sys}^{~i}$ in [K] is the system temperature of the i-th antenna,
$g_i= 2.845 \cdot 10^{-4} \cdot \eta_A \cdot D_m^2$ is the gain in [K/Jy],  
$\eta_A$ is the aperture efficiency, $D_m$ is the antenna diameter in [m], $\eta_c$
is the quantization loss factor, $\Delta \nu$ is the observing bandwidth in [Hz] and $\tau$
is the integration time in [sec].

From the above equation the sensitivity of a VLBI image ($\Delta S$ in [$\mu$Jy/beam]) 
can be estimated using:

\begin{equation}
\frac{1}{\Delta S^2} = \frac{\tau}{T} \cdot \sum_{i,j}^{{\rm all~baselines}} \frac{1}{\sigma_{ij}^2} 
\end{equation}

For the calculations we assume that in addition to the 
existing VLBA (10 x 25\,m antennas), a world array consisting of about $N \simeq 20$ telescopes 
(e.g. the VLBA, the phased VLA, the phased MERLIN interferometer and
some European antennas) also participates in VLBI observations. 
We further assume that observations are made at a frequency of 5\,GHz 
with an observing bandwidth of $\Delta \nu=128$ MHz at present, 
and of $\Delta \nu=512$ MHz in the future (SKA, ARISE),
using 1-bit data sampling ($\eta_c=0.64$), a segmentation time equal to a 
coherence time of $\tau=300$ sec at 5\,GHz and a total integration time on source of $T=12$ h.

Following the specification of SKA outlined in Taylor and Brown (1999), \cite{tb99},
we assume a system temperature $T_{\rm sys}=50$ K, and a total collecting area of 
$A=A_{\rm eff}/\eta_A=2.3 \cdot 10^6$ m$^2$ ($\eta_A=0.5$).
This corresponds to $N=32$ antennas each of $D=300$ m diameter.
We also assume that SKA can be operated either in single antenna mode (some
or all antennas equipped with VLBI recording equipment) or as a phased array,
the latter acting as a single big antenna in a VLBI world array. Since SKA will
be a transit instrument (flat array), the direction-dependent gain (or the aperture efficiency) will be
maximum near zenith. It is desirable that the variations of the gain
as a function of hour angle (or elevation) should not be too strong,
e.g. $\Delta g/g \leq 0.5$ for hour angles of $\pm 6$\,hrs.
For simplicity we neglect any variation of the gain in our calculations.

For comparison of the sensitivity between present VLBI observations
and future observations including SKA, we summarize baseline and
image sensitivities in Table 1. With regard to VLBI observations in 
combination with orbiting antennas (space-VLBI), we included the 8\,m antenna
of the Japanese space mission (VSOP). For a future space-VLBI mission
we used ARISE ($D = 25$ m) as a representative project (e.g.\ Ulvestad and Linfield, 1998, 
\cite{arise}).

\begin{table}
\begin{tabular}{|llll|} \hline
Array        &baseline sensitivity &image sensitivity &comment          \\
             &  [mJy]              &  [$\mu$Jy/beam]  &            \\
VLBA         &  ~2.~~~             &     30.~~        &standard today      \\
World Array  &  ~0.3~~             &     ~5.~~        &best today          \\
SKA  + VLBA  &  ~0.018             &     ~0.4~        &improvement/standard 75\\
SKA(1)-SKA(1)&  ~0.011             &     ~0.04        &improvement/standard 750\\
             &                     &                  &                      \\ \hline                
             &                     &                  &                \\
VLBA(1)-VSOP &  24.~~~ ($\Delta \nu =32$\,MHz)&       &standard today      \\
Bonn-VSOP    &  ~6.~~~             &                  &best today          \\
SKA-ARISE   &  ~0.014 ($\Delta \nu =512$\,MHz)&      &improvement/best 400\\ \hline
\end{tabular}
\caption{Baseline and image sensitivity for present and future VLBI observations. SKA denotes
the phased square kilometer array (32 antennas), SKA(1) denotes a single element ($D=300$ m)
of SKA.}
\end{table}

Table 1 shows that the addition of a phased SKA to an existing global VLBI
network will lower the single-baseline detection threshold by a factor of at least 15, facilitating 
instantaneous fringe visibility detection of compact sources of $\geq 10$\,$\mu$Jy flux density. This 
leads to an improvement of the image sensitivity by a factor of at least 50. For a typical
compact source with a flux density of 1\,Jy the dynamic range in future VLBI images
might then exceed $10^6$, 1--2 orders of magnitude larger than presently achieved.
If SKA were operated as a phased array together with other antennas of the 100\,m
class in a future `world array' (e.g. together with the 100\,m telescopes at Bonn and at Green Bank,
the 300\,m antenna at Arecibo and the Chinese FAST telescope (cf. Peng Bo et al., this conference)),
baseline and image sensitivities of 1 $\mu$Jy and 0.1 $\mu$Jy/beam could be reached.
Since the maximum separation between the individual antennas of the SKA probably will be limited
to $\leq 1000$\,km, the addition of other antennas at larger distances is desirable to
achieve the highest possible angular resolution.

The sensitivity of a two-element interferometer is proportional to the geometric mean
of the collecting areas of the two elements. The combination of a bigger 
with a smaller antenna therefore is equivalent to the use of two medium-size identical antennas.
In a global VLBI array consisting mainly of small- and medium-size antennas, the addition of
one large antenna improves the imaging capabilities of the whole array proportionally
to the number of baselines to the large antenna. This facilitates fringe detection
on weaker sources and, via the construction of closure triangles, fringe detection also
on the less-sensitive baselines (global fringe fitting).

It has been pointed out previously that
the combination of a small orbiting antenna in space with a large antenna on the ground is 
financially much more attractive than launching a big antenna into space. In future space 
VLBI missions (e.g. ARISE), SKA can therefore play a very important role. Its impact on 
space VLBI probably could be even greater than for ground VLBI.
In the lower part of Table 1 we compare the gain of baseline sensitivity in a present
space VLBI experiment with, for example the 100\,m antenna at Bonn and VSOP,
and with SKA and ARISE in the future. The improvement will be two orders of magnitude!
\begin{figure}[t]
\begin{center}
\includegraphics[width=13cm, angle=-90]{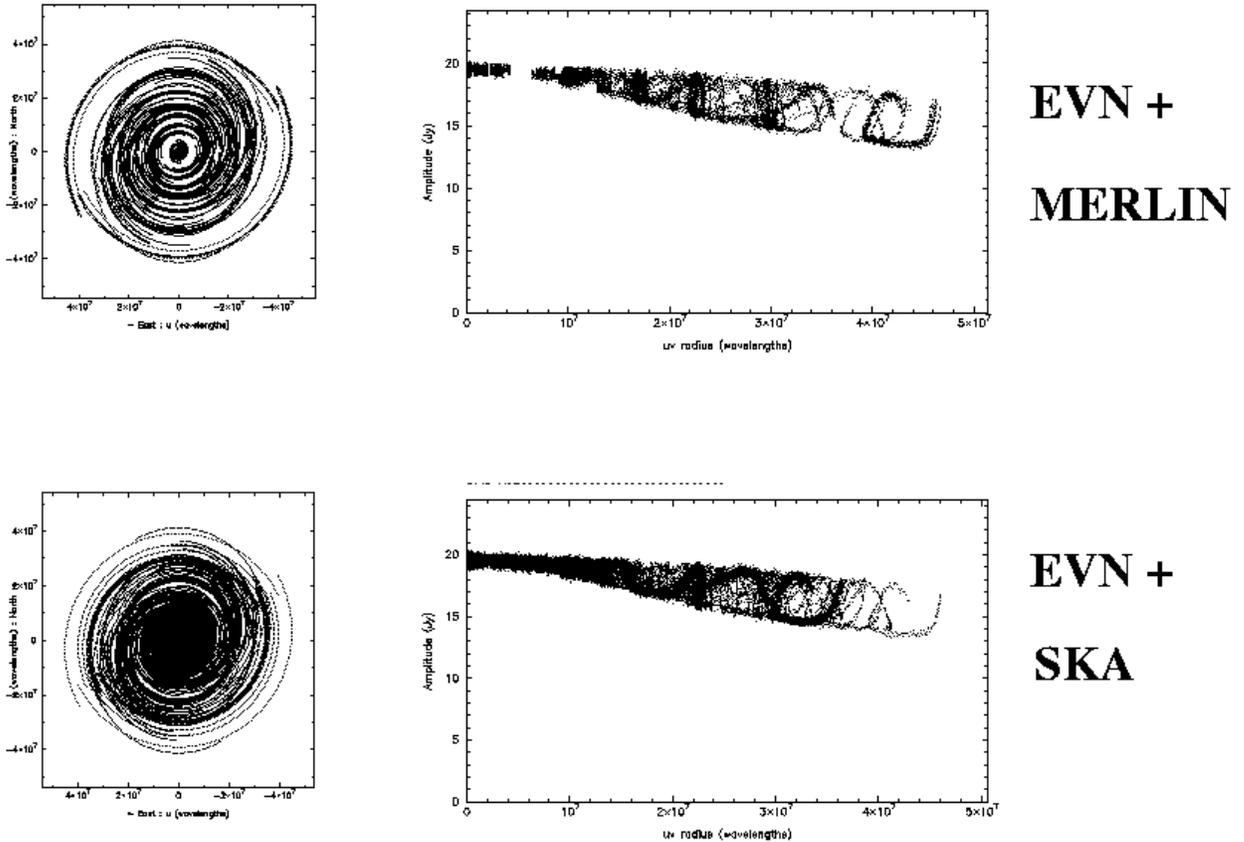}
\caption{Simulation of the uv-coverage for a hypothetical circumpolar source 
observed 12 hours with VLBI. The uv-coverage is shown on the
left. On the right the visibility amplitude is plotted versus
projected baseline length. For the source a core 
dominated core-jet structure of 5 mas length was assumed.
On top a VLBI experiment was simulated
assuming mutual observations of the European VLBI Network (EVN)
and the MERLIN interferometer. In the simulation below, the
MERLIN interferometer was replaced by the SKA interferometer.
For the SKA a geometry as described in the text with an
arm lengths of 500 km was chosen. Note that SKA nearly
fills the uv-plane for $< 15$\,M$\lambda$.
}
\end{center}
\end{figure}

\section{Angular Resolution and uv-Coverage}

The geometrical configuration of the antennas of SKA is still being discussed. For VLBI, a configuration
as proposed by Taylor \& Brown is useful. In Figure 1 we show a simulated uv-coverage
for an experiment involving the European VLBI Network (EVN) and the MERLIN interferometer.
For the simulation we replaced MERLIN by the SKA interferometer. For the configuration of SKA
a Y-shaped geometry (nine antennas arranged on the three 500 km arms) 
and a central area covered by two elliptical rings
(diameter $\leq 50$ km) of concentrically arranged antennas (cf. Taylor \& Braun 1999). 
The 500 km arm length gives an angular resolution of 18 mas.
For a source of $S=1$\,$\mu$Jy, the minimum detectable brightness temperature at 5\,GHz 
will then be $T_B= 150$ K. The addition of the EVN to SKA will improve the angular 
resolution to 4--5\,mas. To reach higher angular resolution, longer baselines have to
be included (unless the antenna spacing of SKA is increased to world-wide separations). 
For VLBI observations with transcontinental baselines, an angular resolution of 1\,mas can be 
achieved in ground-based VLBI observations at 5\,GHz. 
With baselines to an orbiting antenna at 40\,000 km altitude 
(ARISE) 0.3\,mas will become possible. For a source at a redshift of $z=0.1 (1)$, this corresponds 
to a spatial resolution of 0.4 (1.3)\,pc (for \hub). The minimum detectable brightness temperature
in this larger configuration then will be $T_B=10^4$K (ground) and $10^5$K (space),
which is two orders of magnitude lower than presently achieved.

The combination of SKA with existing or future telescope in a global VLBI array 
offers, besides the sensitivity improvement, some other advantages: 

(i) To achieve the highest angular resolution for ground
based observations, the spacing between individual antennas of SKA need not 
be extended over more than a few hundred kilometers. This limits the costs for data links
between the antennas (e.g. fiber optics) and facilitates operation as \underline{one} 
phased array.

(ii) present day VLBI observations usually lack short baselines, rather than long
baselines. To satisfy the interests of non-VLBI observers, SKA has to provide mainly short uv-spacings
(emphasis to 10-100 km baselines). In our simulation (see Figure 1), SKA contributes
mainly at uv-spacings $\leq 15$\,M$\lambda$. In this area
it nearly fills the uv-plane.
Therefore, one of the main advantages SKA could offer for VLBI is the potential 
to close the gap between long and short uv-spacings. This will allow us, better
than presently possible, to image source structures continuously from
milliarcsecond to arcsecond scales (compare with the wide-field imaging technique addressed
by M. Garret, this conference). Future high-fidelity images resulting from 
a combined SKA+VLBI interferometer will allow one to `zoom' continuously
from outer- to inner-scale structures within the same data set. This is  
important for example for the investigation of jets in active galaxies 
(radio galaxies, Seyfert galaxies, etc.) or for the imaging of gravitational lenses. 

(iii) in aperture synthesis, the image quality depends on the number of participating
telescopes. The combination of SKA with other VLBI telescopes would
introduce a large number of baseline combinations ($N(N-1)/2$, $N \geq 30-40$), depending on
how many of the individual SKA antennas will be equipped with 
VLBI recording equipment. This would provide a uv-coverage with high redundancy, 
much more dense than presently synthesized. Facilitated by the usage of closure relations 
(for phases and amplitudes) a much better self-calibration of the data and 
an improved fidelity of the final images (dynamic range: $10^{6...7}$) will be achieved.

\section{General Improvements}

In summary, a next-generation interferometer like SKA will provide for VLBI, high sensitivity ($\mu$Jy-level),
quasi-continuous uv-coverage (from arcseconds to milliarcseconds), a brightness temperature 
limit of $T_B \geq 10^4$ K, and a map fidelity of better than $10^6:1$ (dynamic range).
As a phased array, SKA will provide an angular resolution of 20-30 mas. With other VLBI antennas
added to it, 1 mas can be reached from ground, and about $\sim 0.3$ mas using an orbiting antenna.

With this, first of all, more and fainter compact radio sources can be observed. The study of
more distant quasars (QSO) and radio galaxies (RG) will allow us to understand better their
evolution (luminosity functions) and will allow sophisticated tests on 
cosmology ($H_0$, $q_0$, inflation). Detecting and imaging the large number
of radio quiet QSOs and faint radio counter-parts
of galaxies that are bright at other wavelengths will have its impact on unification models of
AGN. The study
of cores of galaxies and Sgr\,A*-like objects in other galaxies (e.g. M\,81, NGC\,4258)
will eventually lead to detections of supermassive black holes, accretion disks and general-relativistic
effects in their vicinity. In gravitational lenses, the multiple images
can be studied with quasi-continuous resolution from  milliarcseconds to tens of arcseconds,
allowing studies of the cosmological metric (time delay along path) and cosmological
propagation effects. There is hope that even the (dark) lensing galaxies will become
directly observable. With brightness temperatures
$> 10^4$ K, compact extragalactic H\,II regions, regions of starbursts and supernova remnants
would become observable. Systematic studies of supernova expansions in external galaxies can be
used to improve the distance ladder.  With a sensitivity $> 1000$ times better than to day, the 
radio-interferometric investigation of various kinds of radio stars, peculiar binaries, pulsars, etc.\ 
will be pushed forward, revealing insight in other radiation processes than just
thermal or synchrotron radiation. Radio interferometric observations provide the
highest positional accuracy in astrometry (accuracy range: $10^{-3} ... 10^{-6}$ arcsec). 
The detection of planets around $\mu$Jy-bright (radio) stars and the monitoring of stellar
motion should then become feasible for many objects, using phase referencing techniques.

\section{Study of Jets in AGN}

The activity in active galactic nuclei (AGN: QSOs, BL Lacs, Blazars, OVVs, HPQs, ...) manifests
itself primarily in intensity variations (total intensity and polarization) with time scales
ranging from minutes to years. Often, the variations are correlated over wide bands of the electromagnetic
spectrum ($\gamma$-rays to radio wavelengths) and appear first and with shorter duration at
high frequencies and subsequently propagate towards longer wavelengths and longer timescales.
In many sources observed with VLBI, 
correlations between the intensity variations and the ejection of sub- or superluminal
components from the VLBI core are observed.  In the context of such rapid
variability, SKA can contribute significantly to the investigation of the quasar
phenomenon. It could help to answer the important questions of how energy is converted near
black holes, how the highly relativistic jets are made and how they propagate. This
contribution of SKA is complementary to the expected contribution of other large
instruments planned e.g.\ at millimeter wavelengths.

Due to its high observing frequency and angular resolution, VLBI at millimeter wavelengths (mm-VLBI)
can detect new ejected jet plasma (= new components) earliest after the initial
flaring. These features appear at core separations as low as a few ten micro-arcseconds
($40-50$\,$\mu$as are presently achieved at 86\,GHz). In future, the activity in AGN
and their outburst-ejection relations need to be studied in more detail. The future
mm-VLBI networks will include large phased interferometers (e.g. BIMA, IRAM, OVRO, NRO).
The addition of the planned Atacama Large Millimeter-Array (ALMA, the former LSA/MMA)  to mm-VLBI
will yield very high sensitivity (mJy-level). The role ALMA will play for astronomy at
mm-wavelengths (cf. Krichbaum, 1996, \cite{kri96}), SKA will play at cm-wavelengths. 

At the longer mm- and at short cm-wavelengths space-VLBI observations including SKA
will provide high angular resolution, which 
will allow us to follow the injected jet components at or slightly below the self-absorption frequency,
in the transition region between optically-thick and optically-thin radiation of the jet.
Ground based VLBI observations with SKA at the longer cm-wavelengths will provide 
highest sensitivity and will reveal finer details in jets on all angular scales $> 1$\,mas. 
The profiles of jets in total intensity and polarization,
their widths and ridgeline, their curvatures and spectral shapes could be 
imaged continuously from the base (or nozzle) out to the hot spots. For the latter,
their advance speeds could be measured. With a dynamic range of order $10^6$, 
the motion of the more central features in the jet 
(let it be pattern or bulk motion, motion of shocks or
propagation of instabilities) can be followed over much longer distances and times than
presently possible. Loosing a moving and slowly synchrotron-decaying component as it
falls below sensitivity limit
would no longer happen on mas-scales but on scales of tens to hundreds of mas. 
Spectral-index studies will become possible over a larger distance along the jet.

Features in jets usually do not move ballistically. They accelerate or decelerate,
change direction of motion and quite often seem to move along quasi-sinusoidal
bent paths, suggesting motion along spatially bent and helical trajectories.
Before discussing jet intrinsic properties, the geometric effects on 
the observed variability pattern (e.g.\ flux, speed) have to be eliminated.
At present, however, it is unclear whether this helical motion is caused externally,
in a flow along a helical magnetic field e.g.\ anchored in a rotating accretion disk, 
or if it is due to jet intrinsic processes, like Kelvin-Helmholtz or magneto-hydrodynamical
instabilities. With total intensity and polarization
maps of sufficiently high dynamic range it should be possible to discriminate between
intrinsically three-dimensional motion, which would cause systematic and correlated
variations of velocity, brightness and polarization, and alternative kinematical scenarios,
e.g.\ collectively moving patterns, kinks or wiggles, which would be more 
typical for instabilities. From high-quality maps the modes of such instabilities (the
wavelength and amplitude of the oscillation of jet axis and width) could be determined 
much more accurately than at present. This will improve our knowledge of
still largely unknown physical parameters of jets, like intrinsic velocity distribution, 
magnetization, pressure and density contrast (light or heavy jets?).

Another important topic is the physics of shocks and how the
acceleration of particles (electrons, positrons, protons) in jets occurs. 
The relative abundance of these particles in the jets and their energy distribution
is also largely unknown. Obviously, the radiating relativistic particles of the jet 
have to be reaccelerated, but at present it is unclear if this is done only by shocks. 
It is commonly assumed that the components detected with VLBI in jets are related to relativistic 
shocks, which can be moving or stationary, thin or thick, oblique or transverse,
and being polarized parallel or perpendicular to the jet axis. 
The observed flux density variations can be explained by shock-in-jet scenarios, 
at least to some extent.
High dynamic range imaging of jets in total intensity -- and perhaps more important -- 
in polarized light, will allow us to study how the shocks evolve in time and frequency,
if they fill the jet widths or only part of it (filling factor, obliqueness),
if and how they are magnetized (linear or circular polarization), 
how the observed motions couples to the underlying flow (is $\beta_{\rm jet} = \beta_{\rm shock}$ ?), 
and if there is evidence for magnetic reconnection or other collective emission processes.
With higher angular resolution (space-VLBI), the detailed structure of shocks,
their internal stratification, reverse shocks, back-flows and rarefaction regions could be 
investigated and theory could be tested.

\begin{figure}
\begin{center}
\includegraphics[width=11cm]{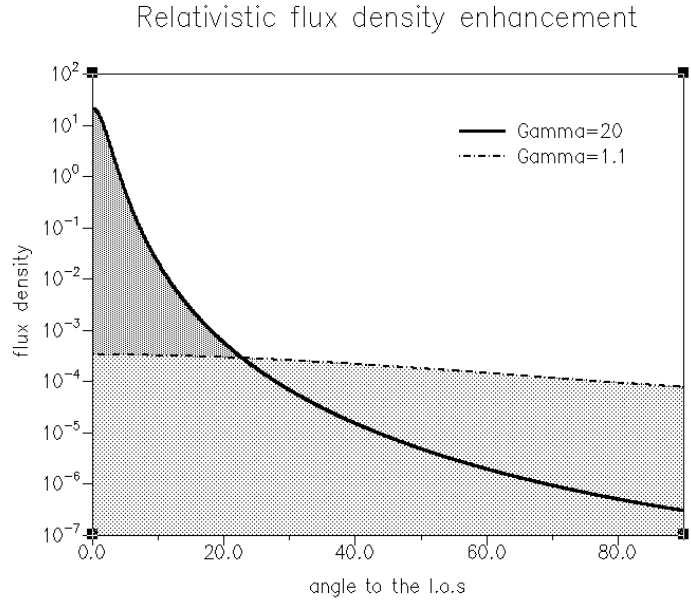}
\caption{Flux density plotted versus inclination angle to the line of
sight for a fast (Lorentz factor $\gamma =20$) and a slow ($\gamma =1.1$) relativistic 
flow. At small angles, the emission from the high-$\gamma$ flow dominates,
while at large angles the low-$\gamma$ flow is brighter. For inclined
jets, this introduces a selection bias; the observer will see the part of the
jet with the appropriate Lorentz factor. To study the internal velocity distribution
in jets of quasars and radio galaxies, high dynamic range images like those
expected from SKA will be needed.
}
\end{center}
\end{figure}

The observed brightness $S$ of a relativistic flow (jet), which is inclined towards the observer
depends on the Doppler-boosting factor $D$, which itself is a function
of Lorentz factor $\gamma = (1- \beta^2)^{-0.5}$ and jet inclination $\theta$:
$S \propto D^{2 + \alpha}$; $D = \gamma^{-1} (1 - \beta \cos{\theta})^{-1}$, where
$\alpha$ is the spectral index (\sprop) and $\beta=v/c$ is the jet speed. 
The relativistic effects determine the intensity
ratio between jet, pointing towards the observer, and counter-jet, pointing
away from him. The extremely high dynamic range ($10^{6...7}$) of future images made with SKA and
VLBI will allow us to measure accurately the jet-to-counterjet ratio even for the so called
one-sided sources (QSOs, BL\,Lacs, ...), which are thought to be oriented at small ($\leq 10-20 \deg$)
inclination angles and in which the brightness of the counter-jet is strongly attenuated. 
For sources that are oriented at larger inclinations (two-sided radio galaxies, Seyfert galaxies, etc.)
the high sensitivity of such images to low-surface-brightness features will make it
possible to overcome the selection bias towards specific jet velocities in inclined
radio jets (see Figure 2). This might lead to the detection of high-velocity flows in sources
oriented at large angles, or low-velocity flows in sources oriented at small angles. 
Such multiple flows are not unexpected and their existence may solve some problems
of present-day models regarding source unification.
From a known jet-to-counter-jet ratio and from the velocities, the intrinsic geometry
and the distance of the sources can be determined. The study of two-sided jets
(cf. A. Roy et al., this conference) is ideally suited to test the unification models,
which try to explain differences between various source classes as due to 
geometrical effects in an inclined system of accretion disk (thin or thick) 
and two sided jet emission (radio galaxies $\leftrightarrow$ blazars, 
Seyfert\,1 $\leftrightarrow$ Seyfert\,2 galaxies, FR\,I galaxies $\leftrightarrow$ BL\,Lacs, etc.).
In such a geometry the separation between the footpoints of jet and counter-jet
depends on observing frequency and on the brightness of the counter-jet, which is reduced
by absorption from the accretion disc located in front of it (free-free absorption).
High angular resolution studies with VLBI and SKA in spectroscopic and continuum
mode at different frequencies will allow us to determine the geometrical and physical
parameters of accreting supermassive black hole systems and their jets. This will help us
to unravel the relation of the various classes of AGN (quasars, galaxies) to each other,
improving our knowledge of the cosmological evolution of galaxies and quasars.

\section{Summary}

Ground-based observations are logistically easier to perform, can be used by a 
larger scientific community for a longer period of time, and usually are cheaper than pure 
satellite-based
experiments. One of the two windows that the atmosphere offers to observers
is the radio window. SKA offers micro-Jansky sensitivity for research in this band. 
With this sensitivity an unpredictable number of discoveries can be made, 
even though the radio band has been exploited for more than 50 years. 
With a collecting area of order of a million square meters, SKA can provide a 
break-through 
to imaging in radio astronomy and to our knowledge about the Universe. 
The construction of SKA will be technically and financially challenging, but worth
the effort. 

In this paper, we discussed the addition of SKA as a 
phased array to existing or future VLBI networks. This will allow 
studies of virtually all types of compact galactic and extragalactic radio sources 
with unrivaled image quality and angular resolution. For the first time it will be
possible to `zoom' into extended radio sources with continuous uv-coverage ranging from
arcseconds to milliarseconds. In conjunction with orbiting antennas,
an instrument like SKA will be required to provide the sensitivity needed for 
studies of the most distant (and therefore faint) radio sources with sub-milliarcsecond resolution.
For closer sources, the high spatial resolution which VLBI with SKA offers will facilitate
finer images and more detailed studies of the central regions of AGN and their jets.
In future interferometers operating at millimeter wavelengths (e.g.\ ALMA, mJy-sensitivity) and 
centimeter wavelengths (e.g.\ SKA, $\mu$\,Jy-sensitivity) will complement each other perfectly.
The complementary scientific return of both instruments will allow to answer many of the
unsolved problems and questions, in particular those related to the quasar phenomenon,
to the cosmological evolution of galaxies and to the chemistry of matter in space.

\section*{Acknowledgements}

For their contributions and stimulating discussions we thank Drs. 
Andrew Lobanov, Alex Kraus, Alan Roy, and Eduardo Ros.
\end{document}